\def\dir{./}
\documentclass[useAMS,usenatbib]{mn2e}
\usepackage[fleqn]{amsmath}
\usepackage{amssymb}
\usepackage[super]{nth}
\usepackage{natbib}
\usepackage{graphicx}
\usepackage{float}
\usepackage{mathrsfs}
\usepackage{mathtools}
\usepackage{multirow}
\usepackage[usenames,dvipsnames]{color}
\usepackage{\dir aas_macros}
\usepackage{comment}
\usepackage{subfigure}
\usepackage{hyperref}
\usepackage{mathrsfs}
\usepackage{epstopdf, dcolumn}
\DeclareGraphicsExtensions{.pdf}
\usepackage{wasysym}
\usepackage{dashrule}
\usepackage{xcolor}
\usepackage{arydshln}
\usepackage{threeparttable}
\usepackage[export]{adjustbox}
\usepackage[figuresleft]{rotating}
\usepackage{array,multirow}

\newlength\replength
\newcommand\repfrac{.33}

\newcommand\rulewidth{.6pt}
\newcommand\tdashfill[1][\repfrac]{\cleaders\hbox to \replength{%
		\smash{\rule[\arraystretch\ht\strutbox]{\repfrac\replength}{\rulewidth}}}\hfill}

\newcommand\tdotfill[1][\repfrac]{\cleaders\hbox to \replength{%
		\smash{\raisebox{\arraystretch\dimexpr\ht\strutbox-.1ex\relax}{.}}}\hfill}

\newcommand{\appropto}{\mathrel{\vcentre{
			\offinterlineskip\halign{\hfil$##$\cr
				\propto\cr\noalign{\kern2pt}\sim\cr\noalign{\kern-2pt}}}}}

\newcommand\lsim{\mathrel{\rlap{\lower4pt\hbox{\hskip1pt$\sim$}}
        \raise1pt\hbox{$<$}}}
\newcommand\gsim{\mathrel{\rlap{\lower4pt\hbox{\hskip1pt$\sim$}}
        \raise1pt\hbox{$>$}}}
\newcommand{\Rom}[1]{\uppercase\expandafter{\romannumeral #1}}
\newcommand{\rom}[1]{\lowercase\expandafter{\romannumeral #1}}

\newcommand{\hone}{\mathrm{H}\textsc{i}}
\newcommand{\hii}{\mathrm{H}\textsc{ii}}
\newcommand{\tocm}{{21\textsc{cmFAST}}}
\newcommand{\cmmc}{21\textsc{CMMC}}

\newcommand{\msol}{{\rm M}_\odot}

\defcitealias{Park2019MNRAS.484..933P}{P19}

\begin{document}

\title[Reionization inference from the CMB]{Reionization inference from the CMB optical depth and E-mode polarization power spectra}
\author[Qin et al.]{Yuxiang Qin$^{1}$\thanks{E-mail: Yuxiang.L.Qin@gmail.com}, Vivian Poulin$^2$, Andrei Mesinger$^1$, Bradley Greig$^{3,4}$
	\newauthor Steven Murray$^{5}$ and Jaehong Park$^1$\\
	$^{1}$Scuola Normale Superiore, Piazza dei Cavalieri 7, I-56126 Pisa, Italy\\
	$^{2}$Laboratoire Univers \& Particules de Montpellier, CNRS, Universit$\acute{e}$ de Montpellier, Place Eug$\grave{e}$ne Bataillon, 34095 \\\ \  Montpellier Cedex 05, France\\
	$^{3}$School of Physics, University of Melbourne, Parkville, VIC 3010, Australia\\
	$^{4}$ARC Centre of Excellence for All Sky Astrophysics in 3 Dimensions (ASTRO 3D)\\
	$^{5}$School of Earth and Space Exploration, Arizona State University, Tempe, AZ, USA
}
\maketitle
\label{firstpage}

\begin{abstract}
The Epoch of Reionization (EoR) depends on the complex astrophysics governing the birth and evolution of the first galaxies and structures in the intergalactic medium.
EoR models rely on cosmic microwave background (CMB) observations, and in particular the large-scale E-mode polarization power spectra (EE PS), to help constrain their highly uncertain parameters.  However, rather than directly forward-modelling the EE PS, most EoR models are constrained using a summary statistic -- the Thompson scattering optical depth, $\tau_e$.  Compressing CMB observations to $\tau_e$ requires adopting a basis set for the EoR history.  
The common choice is the unphysical, redshift-symmetric hyperbolic tangent ({\it Tanh}) function, which differs in shape from physical EoR models based on hierarchical structure formation.
Combining public EoR and CMB codes, {\tocm} and {\textsc CLASS}, here we quantify how inference using the $\tau_e$ summary statistic
impacts the resulting constraints on galaxy properties and EoR histories.  Using the last {\it Planck} 2018 data release, we show that the marginalized constraints on the EoR history are more sensitive to the choice of the basis set ({\it Tanh} vs physical model) than to the CMB likelihood statistic ($\tau_e$ vs PS).
For example, EoR histories implied by the growth of structure show a small tail of partial 
reionization extending to higher redshifts.
However, biases in inference using $\tau_e$ are negligible for the {\it Planck} 2018 data.
{\color{black}Using EoR constraints from high-redshift observations including the quasar dark 
fraction, galaxy UV luminosity functions and CMB EE PS, our physical model recovers $\tau_e = 
0.0569_{-0.0066}^{+0.0081} $.}
\end{abstract}


\section{Introduction}

The epoch of reionization (EoR) leaves footprints in the observed cosmic microwave background (CMB) as the photons Thomson scatter off free electrons.
These include damping the primary temperature anisotropies, 
inducing secondary anisotropies from the bulk motion of ionized gas
(i.e. the kinetic Sunyaev–Zel'dovich effect), 
and prompting curl-less (i.e. E-mode) polarization from the CMB quadrupole (e.g. \citealt{Sunyaev1980MNRAS.190..413S,Vishniac1987ApJ...322..597V,Hu2000ApJ...529...12H,Hu2002ARA&A..40..171H,McQuinn05,Dvorkin2009PhRvD..79j7302D}).
Of these, the large-scale E-mode polarization anisotropies are a particularly powerful probe of the EoR as they are less plagued by degeneracies  and systematics (e.g. \citealt{Reichardt_2016}).
Reionization models can therefore constrain their largely uncertain parameters that describe the ionizing emissivity of the early Universe, through forward-modelling the EE autocorrelation power spectra (PS) and comparing against measurements from the {\it Planck} satellite (\citealt{Planck2016A&A...596A.108P,Planck2019arXiv190712875P}; e.g. \citealt{Hu2003PhRvD..68b3001H, Mortonson2008ApJ...672..737M, Miranda2017MNRAS.467.4050M, Hazra19}).

Nevertheless, most reionization models do not directly use the CMB PS to constrain their parameters.
Instead, they use a summary statistic which has become one of the de-facto standard cosmological parameters -- the direction-averaged Thomson scattering optical depth, $\tau_e$.
In going from the observed PS to $\tau_e$, one needs to adopt a basis set for the EoR history -- a parametrization of the redshift evolution of the comoving number density of free electrons ($n_{e}$).
Early works adopted a simple step function reionization at a given redshift $z_{\rm re}$ \citep{Page2007ApJS..170..335P,Spergel2007ApJS..170..377S}.  Currently, the most common choice has a hyperbolic tangent functional form ({\it Tanh}; \citealt{Lewis_2008}) parametrized by a reionization midpoint ($z_{\rm re}$)  and a redshift duration ($\Delta_{\rm re}$; see equation \ref{eq:tanh}).
For example, the latest constraints on $\tau_e$
published by {\it Planck}, $\tau_e = 0.0522\pm0.0080$ (the {\it TT+lowE} reconstruction in \citealt{Planck2018arXiv180706209P}), were generated by fixing a width of $\Delta_{\rm re}=0.5$,  sampling a flat prior over $z_{\rm re}$, and comparing against the observed large-scale E-mode PS (and also the temperature autocorrelation).

However, the redshift-symmetric evolution given by {\it Tanh}
differs in shape from both physical and empirical models of EoR history, based on the growth of cosmic structure and/or fit to observed galaxy luminosity functions (LFs; e.g. \citealt{Choudhury2006MNRAS.371L..55C,Raskutti2012,Koh2016,Greig2017,Qin2017a,Kulkarni2018,Gorce2018A&A...616A.113G,Roy2018JCAP...05..014R,Qin2020arXiv200304442Q}).
Therefore, {\it computing a likelihood using the $\tau_e$ summary statistic instead of directly forward-modelling the CMB PS can bias EoR model constraints} (e.g. \citealt{Douspis2015A&A...580L...4D,Planck2016A&A...596A.108P,Miranda2017MNRAS.467.4050M, Hazra19}).
For example, \citet{Miranda2017MNRAS.467.4050M} and \citet{Heinrich_2018} claimed (though see \citealt{Millea2018A&A...617A..96M} and \citealt{Planck2018arXiv180706209P}) that the {\it Planck} 2015 E-mode PS prefers reionization histories with an extended tail of partial reionization towards very high redshifts ($z{>}15$).  This would have significant implications for our understanding of the very first galaxies.  However, quantifying such claims is difficult without directly performing inference on galaxy model parameters.

In this work, we use a physically-motivated EoR model to quantify how inference using the $\tau_e$ summary statistic (instead of directly forward-modelling the E-mode PS)
impacts the resulting constraints on galaxy properties and EoR histories.  Using the last {\it Planck} data release, we show that the marginalized constraints on the EoR history are far more sensitive to the choice of the basis set ({\it Tanh} vs physical model) than to the CMB likelihood statistic ($\tau_e$ vs PS).
Specifically, we use the latest v3.0.0 release\footnote{\url{https://github.com/21cmfast/21cmFAST}} 
(Murray et al. in prep.) of {\tocm} \citep{Mesinger2007ApJ...669..663M,Mesinger2011MNRAS.411..955M}, whose parametrization for galaxy properties is informed by high-redshift UV LFs \citep{Park2019MNRAS.484..933P}.
To calculate the CMB PS for a given reionization history and compute the corresponding likelihood,
we add the Cosmic Linear Anisotropy Solving System 
(\textsc{class}\footnote{\url{https://github.com/lesgourg/class_public}}; 
\citealt{Lesgourgues2011arXiv1104.2932L}) Boltzmann solver and the {\it Planck} 
likelihood codes (\textsc{plc}\footnote{\url{http://pla.esac.esa.int/pla}}; \citealt{Planck2016A&A...594A..11P,Planck2019arXiv190712875P}) to the upcoming 
v1.0.0 release\footnote{\url{https://github.com/21cmfast/21CMMC}} of the public {\cmmc} 
Markov chain Monte Carlo (MCMC) framework \citep{Greig2015MNRAS.449.4246G,Greig2017MNRAS.472.2651G}.
All codes developed here are publicly available.

This paper is organized as follows. We present our analysis of the {\it Planck} data and 
discuss the difference between the 2015 and 2018 results in Section \ref{sec:model_cmb}.
In Section \ref{sec:model_eor}, we briefly introduce our EoR model (Sec. \ref{subsec:21cmfast})
and show the resulting constraints inferred from CMB and 
other observations.  We quantify the bias from the choice of basis set for EoR histories in Sec. \ref{subsec:tanhvs21cmfast} and the choice of likelihood statistics in Sec. \ref{subsec:tauvsps}.  We summarize our results and conclusions in Section \ref{sec:conclusion}.

\section{Modelling CMB observables}\label{sec:model_cmb}

\textsc{class} \citep{Lesgourgues2011arXiv1104.2932L} computes
	CMB anisotropies including temperature and polarization, and calculates their autocorrelation and cross-correlation
	PS. In its default configuration,
	\textsc{class} computes the ionization history from $z{=}10^4$ and
	throughout recombination (e.g. via the code \textsc{recfast}; 
	\citealt{Seager1999ApJ...523L...1S,Chluba2010MNRAS.402.1195C}), and includes a parametrized function for the EoR history.
	The EoR is assumed to have the {\it Tanh} form as follows
	\begin{equation}\label{eq:tanh}
	n_e {=} \frac{n_{\rm H}{+}n_{\rm He}}{2} \left\{1{+}\tanh\left[\left(1{-}\left(\frac{1{+}z}{1{+}z_{\rm re}}\right)^{1.5}\right)\frac{1{+}z_{\rm re}}{1.5\Delta_{\rm re}}\right]\right\}
	\end{equation}
	with ${n}_{\rm H}$ and ${n}_{\rm He}$ representing the average comoving number density of hydrogen and helium, respectively.
	Instead of using the default {\it Tanh} parametrization,
	here we forward-model the CMB observables by passing 
	any given ionization history directly to \textsc{class}.
	We then compare the theoretical PS against observations using the {\it Planck} likelihood codes (\citealt{Planck2016A&A...594A..11P,Planck2019arXiv190712875P}; see more in Table \ref{tab:planck}).
	On the other hand,	the integrated history of reionization 
	can also be summarized using the Thomson scattering optical depth
	\begin{equation}\label{eq:tau_e}
	\tau_e = \int_{0}^{z_{\rm d}} c H^{-1}\left(1{+}z\right)^2 n_e\sigma_{\rm T},
	\end{equation}
	where $z_{\rm d}{\sim}1100$, $c$, $H(z)$ and $\sigma_{\rm T}{=}6.65\times10^{-25}{\rm cm^2}$ are the 
	redshift of the last scattering surface at recombination,
	the speed of light, 
	the Hubble constant at redshift $z$, and
	Thomson scattering cross-section,  
	respectively. Below we also 
	perform EoR inference with a Gaussian likelihood ($\mathscr{L}$) computed using $\tau_e$ from {\it Planck} \citep{Planck2016A&A...594A..16P,Planck2018arXiv180706209P}. This allows us to compare the resulting EoR parameter constraints to those obtained from using directly the EE PS for the likelihood.

\subsection{Preference for an early reionization?}
The mapping of CMB PS to the $\tau_e$ summary statistic depends on the chosen basis set and corresponding priors for the EoR history.  The sensitivity of the resulting $\tau_e$ constraints to this choice has been debated extensively in the literature (e.g. \citealt{Douspis2015A&A...580L...4D, Planck2016A&A...596A.108P, Heinrich2016,Reichardt_2016,Planck2018arXiv180706209P, Millea2018A&A...617A..96M, Hazra19}).
As mentioned in the introduction, using different EoR basis sets to infer $\tau_e$ from the CMB PS  and then from $\tau_e$ to astrophysical parameters could result in biases.

Constraining astrophysics directly using the CMB PS bypasses this issue.  For instance,
\citet{Miranda2017MNRAS.467.4050M} adopted a simple EoR galaxy model, parametrized by the ionizing efficiencies of Pop-II or Pop-III dominated galaxies, and constrained their model parameters directly from the {\it Planck} 2015 \citep{Planck2016A&A...596A.108P} PS observations (see also \citealt{Hazra19}).
The top panel of Fig. \ref{fig:miranda} shows their best-fit models, where
\begin{enumerate}
	\item \textit{Pop-II} only considers UV ionizing photons from Pop-II star-dominated galaxies;
	\item \textit{Pop-III} considers UV ionizing photons from both Pop-II and Pop-III star-dominated galaxies; and
	\item \textit{Pop-III, self-regulated} is similar to the previous model but assumes a significant 
	contribution from Pop-III stars in the early universe ($z{\sim20}$) before their formation 
	becomes completely quenched when $n_e/n_{\rm H}$ reaches 0.2.
\end{enumerate}
Also shown is a \textit{Tanh} model that assumes a fairly sharp transition at 
$z_{\rm re}{\sim}10$ (see equation \ref{eq:tanh}). Comparing the corresponding large-scale E-mode polarization PS of these 
models against the {\it Planck} 2015 measurement \citep{Planck2016A&A...596A.108P}, 
\citet{Miranda2017MNRAS.467.4050M} concluded {\it Planck} 2015 might favour a significant 
UV ionizing photon contribution from Pop-III star-dominated galaxies in the early Universe and
the resulting optical depth is much higher than 
what the default {\it Tanh} parametrization suggests.
This example illustrates how a reionization model informed only by the CMB optical depth could result in biased constraints, compared to using the PS directly in the inference.

\subsubsection{From Planck 2015 to 2018}\label{subsubsec:planck}

From their 2015 to 2018 data release, the {\it Planck} collaboration has made 
tremendous efforts in improving the characterization and removal of systematic uncertainties 
affecting the polarization data of the {\it Planck} High Frequency Instrument (HFI) on large angular 
scales \citep{Planck2016A&A...596A.107P}. With such improvements, the {\it Planck} collaboration has 
shown that the mean value and uncertainties of the optical depth deduced from the low-$\ell$ 
data assuming the {\it Tanh} EoR model have significantly decreased from $\tau =0.078 \pm 0.010$ (the {\it TT+lowP} reconstruction in \citealt{Planck2016A&A...594A..13P}) 
to $\tau=0.0522\pm0.0080$ ({\it TT+lowE} in \citealt{Planck2018arXiv180706209P}). 

Before discussing the difference between EoR inference from the CMB optical depth 
and PS, we revisit the 4 best-fit models from \citet{Miranda2017MNRAS.467.4050M} mentioned above,
using the updated measurement from {\it Planck} 2018 \citep{Planck2019arXiv190712875P} to 
see if the latest data still supports an earlier reionization and larger optical depths.

\begin{figure}
	\begin{minipage}{\columnwidth}
		\begin{center}
			\includegraphics[width=0.97\textwidth]{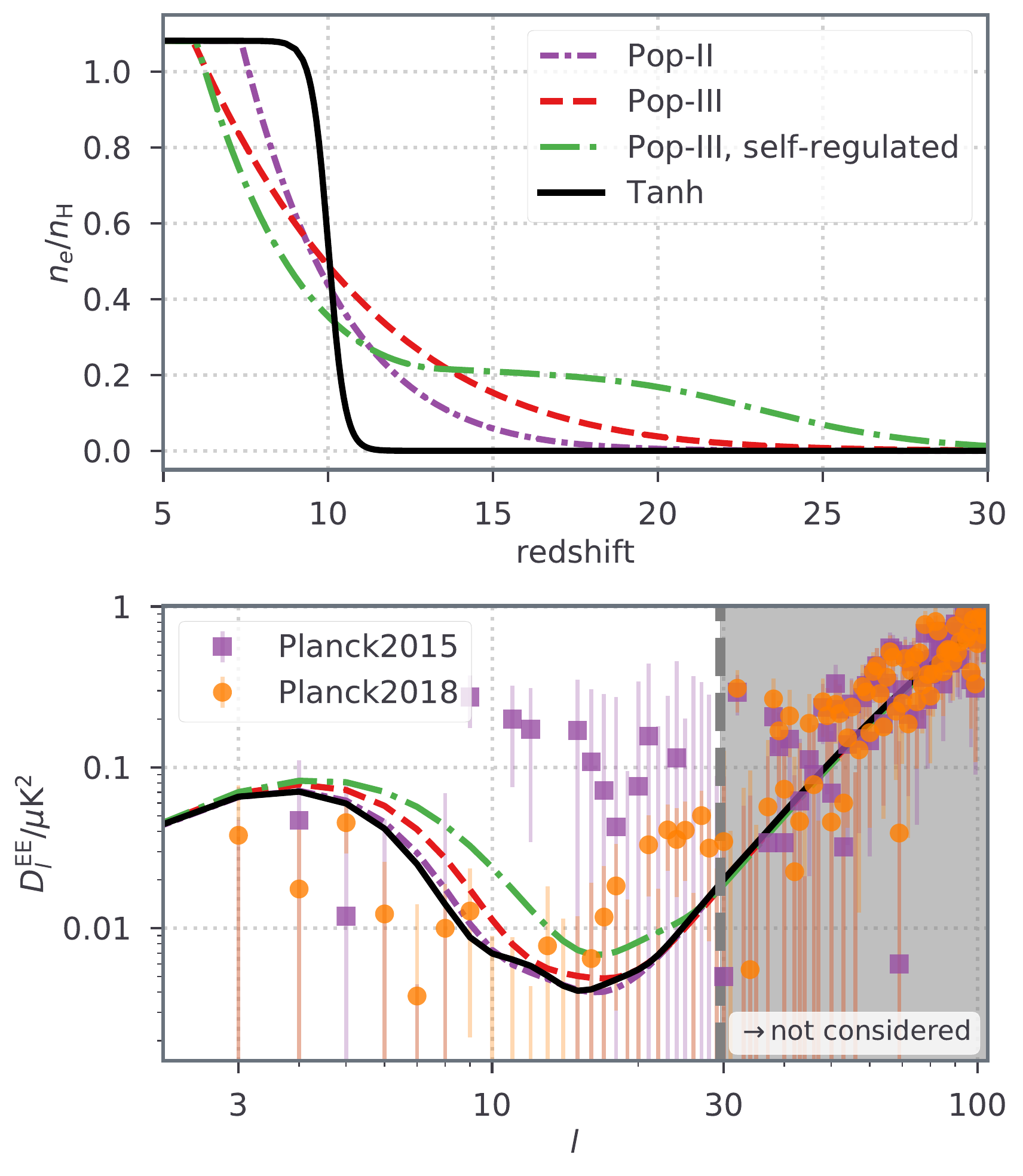}\vspace*{-2mm}
		\end{center}
	\end{minipage}
	\caption{\label{fig:miranda} Reanalysing models from \citet{Miranda2017MNRAS.467.4050M} 
		with cosmological parameters given by the {\it TT+lowP} reconstruction in 
		\citet{Planck2018arXiv180706209P}. \textit{Top panel:} the best-fit EoR models from 
		\citet{Miranda2017MNRAS.467.4050M}. \textit{Bottom panel:} corresponding PS of 
		the E-mode polarization anisotropies with {\it Planck} 2015 and 2018 measurements 
		shown as squares and circles, respectively.  Only the large-scale ($2{\le} \ell{\le}29$) PS is considered when 
		evaluating the likelihoods as these scales are most relevant for the EoR.}
\end{figure}
\begin{table}
	\caption{Revisiting the models from \citet{Miranda2017MNRAS.467.4050M}, using both {\it Planck} 2015 and 2018 data. The models correspond to those shown in Fig. \ref{fig:miranda}; the parameters are fixed and are not allowed to vary so as to better fit the 2018 data.  For each model, we list the optical depth and the difference in the reduced $\chi^2$ with respect to the {\it Pop-II} model (computed using the {\it Planck} likelihood code).
}
	\vspace{-3mm}
	\begin{threeparttable}
		\label{tab:planck}
		\begin{tabular}{l|cc|cc}
			\hline \hline
			\multirow{2}{*}{Model} & \multicolumn{2}{c|}{{\it Planck} 2015\tnote{a}}& \multicolumn{2}{c}{{\it Planck} 2018\tnote{b}}\\\cline{2-5}
			& $\tau_e$&$\chi^2-\chi^2_{\rm PopII}$& $\tau_e$\tnote{c}&$\chi^2-\chi^2_{\rm PopII}$\\\hline
			Tanh&0.0792&1.13&0.0788&-1.14\\
			Pop-II\tnote{d}&0.0832&0&0.0827&0\\
			Pop-III&0.0926&-0.96&0.0921&5.47\\
			Pop-III,self-&\multirow{2}{*}{0.1049}&\multirow{2}{*}{-2.27}&\multirow{2}{*}{0.1043}&\multirow{2}{*}{16.10}\\
			regulated&&&&\\
			\hline
		\end{tabular}
		\begin{tablenotes}
			\item[a] This column assumes cosmological parameters (i.e. the density and Hubble constant) from the {\it TT+lowP} reconstruction in \citet{Planck2016A&A...594A..13P} and calculates $\chi^2$ using the {\it \scriptsize low\_l/bflike/lowl\_SMW\_70\_dx11d\_2014\_10\_03\_v5c\_Ap} likelihood in \textsc{plc} 2.0 \citep{Planck2016A&A...594A..11P}. This likelihood considers both EE, TT, TE and BB components.
			\item[b] Results in this column, similarly to the {\it Planck} 2015 column, assume cosmological parameters from the {\it TT+lowE} reconstruction in \citet{Planck2018arXiv180706209P}, and use the {\it \scriptsize low\_l/simall/simall\_100x143\_offlike5\_EE\_Aplanck\_B} likelihood in \textsc{plc} 3.0 \citep{Planck2019arXiv190712875P}. This likelihood allows users to consider only the EE PS, which is used in this work.
			\item[c] For each model, there are slight changes in the resulting optical depth when comparing the {\it Planck} 2015 and 2018 columns. This is due to the variation in the Hubble constant (see equation \ref{eq:tau_e}) when different cosmological parameters are adopted.
			\item[d] $\chi^2_{\rm PopII} {=} 10492.45$ (407.50) for {\it Planck} 2015 (2018).
		\end{tablenotes}
	\end{threeparttable}
\end{table}

We show the PS of the E-mode polarization anisotropies using the
\citet{Miranda2017MNRAS.467.4050M} EoR models in the bottom panel of 
Fig. \ref{fig:miranda} assuming {\it Planck} 2018 {\it TT+lowE} cosmology
($\Omega_{\mathrm{m}}, \Omega_{\mathrm{b}}, \Omega_{\mathrm{\Lambda}}, h, \sigma_8$ = 
0.321, 0.04952, 0.679, 0.6688, 0.8118; \citealt{Planck2018arXiv180706209P}). We present the 
relative $\chi^2{\equiv}-2\ln \mathscr{L}$ and the corresponding optical depth\footnote{Note that the updated optical depths from the 4 models presented in
		Fig. \ref{fig:miranda} using {\it Planck} 2018 cosmology are all higher than the reported
		value ($0.0522\pm0.0080$; \citealt{Planck2018arXiv180706209P}). This is due to the fact that the model parameters were chosen to fit the 2015 data, not the 2018. Here we do not vary the parameters of the
		\citet{Miranda2017MNRAS.467.4050M} models but instead use a different parametrization introduced below.} in Table \ref{tab:planck}, together
with the results using {\it Planck} 2015 {\it TT+lowP} cosmology 
($\Omega_{\mathrm{m}}, \Omega_{\mathrm{b}}, \Omega_{\mathrm{\Lambda}}, h, \sigma_8$ = 
0.315, 0.04904, 0.685, 0.6731, 0.829; \citealt{Planck2016A&A...594A..13P}) for 
comparison. We conclude that, unlike {\it Planck} 2015, the 
2018 measurement no longer prefers an earlier reionization or a significant contribution from 
Pop-III stars to early reionization -- the likelihood decreases when ionization
starts at earlier times (see also, e.g. \citealt{Millea2018A&A...617A..96M}).
This is mainly driven by the reduced amplitude at multipole 
$\ell{\sim}10 - 20$ in the updated {\it Planck} E-mode PS (see the bottom panel of Fig. \ref{fig:miranda}).
Consequently, models with a better fit to the 
updated PS measurement also return an optical depth closer to the reported value from 
\citet{Planck2018arXiv180706209P}, indicating that EoR inference biases
from using the CMB optical depth might be insignificant in {\it Planck} 2018 (see also v2 of \citealt{Hazra19}).

For the remainder of the paper, we focus only on the {\it Planck} 2018 data.  Using a flexible, physical EoR model, we forward-model the EE PS, and quantify the bias from using the $\tau_e$ summary statistic instead of the PS.

\section{EoR inference from the CMB}\label{sec:model_eor}

To compute the impact of realistic reionization histories on the CMB PS, 
we connect {\tocm} (\citealt{Mesinger2007ApJ...669..663M,Mesinger2011MNRAS.411..955M}; Murray et al. in prep) 
to \textsc{class} \citep{Lesgourgues2011arXiv1104.2932L}.
Specifically, for a given set of cosmological and astrophysical parameters, {\tocm} performs a 3D EoR simulation\footnote{We do not explicitly model helium reionization. Instead we assume 
helium to be singly ionized following the same rate as hydrogen
before becoming fully ionized at $z{=}3$
\citep{Hogan1997AJ....113.1495H,Worseck2011ApJ...733L..24W,Worseck2016ApJ...825..144W}.}.  The volume-averaged EoR history of that simulation is passed to \textsc{class}, which then computes the corresponding CMB PS.

 With this interface, we use the MCMC driver {\cmmc} \citep{Greig2015MNRAS.449.4246G, Greig2017MNRAS.472.2651G}
to sample two EoR models: (i) the astrophysical galaxy-based parametrization native to {\tocm} (see below); and (ii) the commonly used, two-parameter {\it Tanh} model (see equation \ref{eq:tanh}).  In addition to varying the EoR parameterization, we also check how the inference is affected by the choice of likelihood statistic: (i) $\tau_e$; or (ii) the EE PS.  Specifically, for (i) we use $\tau_e=0.0522\pm0.0080$ from the {\it TT+lowE} result in \citet{Planck2018arXiv180706209P}; for (ii) we use the large-scale ($2{\le} \ell {\le} 29$) E-mode polarization measurements and the likelihood from \citet{Planck2019arXiv190712875P}.

\subsection{Modelling the EoR}\label{subsec:21cmfast}
Our EoR model describes galaxy properties mostly using power-law scaling relations with respect 
to their host halo masses \citep{Park2019MNRAS.484..933P} and calculates the 3D reionization evolution following an excursion-set approach based on the cumulative number density of ionizing photons and recombinations (e.g. \citealt{Furlanetto2004ApJ...613....1F, Sobacchi2014MNRAS.440.1662S}).
We 
introduce some basic characteristics of our model below and refer interested readers 
to the aforementioned references for more details.

We start from an initial Gaussian realization of the density and velocity fields 
\citep{Mesinger2007ApJ...669..663M} in a large-volume (250Mpc), grid-based, 
high-resolution (${\sim}$0.65Mpc; i.e 250Mpc/384) simulation box assuming periodic 
boundary conditions. These fields are then evolved according to second-order Lagrangian 
perturbation theory \citep{Scoccimarro1998MNRAS.299.1097S}, and re-gridded to a lower resolution 
(${\sim}$1.95Mpc; i.e. 250Mpc/128) for the sake of computing efficiency. 
Then, for each cell centred at a spatial position and redshift of $({\bf r}, z)$,
we compare the cumulative number per baryon of ionizing photons\footnote{In this work, we do not consider ionization by X-ray photons or earlier objects such as minihalo-hosted galaxies, and their corresponding parameters.  Although efficient at heating the IGM before reionization, X-rays and minihalos are expected to have a very minor contribution to the EoR for reasonable galaxy models (e.g. \citealt{RO04, McQuinn12, Mesinger2013MNRAS.431..621M, Ross17, Eide18,Qin2020arXiv200304442Q}).} ($\bar{n}_{\rm ion}$) to that of recombinations ($\bar{n}_{\rm rec}$) in spheres with decreasing radii \citep{Furlanetto2004ApJ...613....1F, Sobacchi2014MNRAS.440.1662S}. A cell is ionized if at any radius $R$,
\begin{equation}\label{eq:ionization}
\bar{n}_{\rm ion} \ge (1+\bar{n}_{\rm rec}).
\end{equation}
Unresolved $\hii$ regions (smaller than the cell size) are accounted for according to \citet{Zahn2011MNRAS.414..727Z}.

The cumulative number of ionizing photons per baryon is obtained with
\begin{equation}\label{eq:n_ion}
\bar{n}_{\rm ion}\left({\bf r},z|R,\delta_{\rm R|_{{\bf r},z}}\right) {=}  \int {\rm d}M_{\rm vir} {\phi}f_{\rm duty} f_* \frac{\Omega_{\mathrm{b}}}{\Omega_{\mathrm{m}}}\frac{M_{\rm vir}}{\rho_{\rm b}}n_{\gamma} f_{\rm esc}
\end{equation}
where $\delta_{\rm R|_{{\bf r},z}}{\equiv}\rho_{\rm b}/\bar{\rho}_{\rm b} {-}1 $ is the 
average overdensity within the spherical region, $\rho_{\rm b}$ 
and $\bar{\rho}_{\rm b}$ represent the baryonic density and its cosmic mean. In equation (\ref{eq:n_ion})
\begin{enumerate}
	\item $M_{\rm vir}$ and 
$\phi\left(M_{\rm vir},z|R, \delta_{\rm R|_{{\bf r},z}}\right)$ are the halo mass and halo 
mass function; 
	\item $f_{\rm duty}\left(M_{\rm vir}\right) {=} \exp\left(-{M_{\rm turn}}/{M_{\rm vir}}\right)$, with a characteristic mass ($M_{\rm turn}$) as a free parameter, accounts for a decreasing occupation fraction of star forming galaxies inside smaller halos due to inefficient cooling, photo-heating feedback 
\citep{Efstathiou1992MNRAS.256P..43E,Shapiro1994ApJ...427...25S,Thoul1996ApJ...465..608T,Hui1997MNRAS.292...27H,Sobacchi2013MNRAS.432L..51S,Sobacchi2014MNRAS.440.1662S} or supernovae feedback \citep{Hopkins2014MNRAS.445..581H, Hopkins2017, Wyithe2013MNRAS.428.2741W,Sun2015,Mutch2016};
 \item $f_*\left(M_{\rm vir}\right) {=} \min\left[1, f_{*,10} \left({M_{\rm vir}}/{10^{10}{\msol}}\right)^{\alpha_*}\right]$ is the fraction of galactic gas in stars and is assumed to scale with the host halo mass 
\citep{Moster2013MNRAS.428.3121M,Sun2015,Mutch2016,Tacchella2018ApJ...868...92T,Behroozi2019MNRAS.488.3143B,Yung2019MNRAS.490.2855Y} 
according to the two free parameters, $f_{*,10}$ and $\alpha_*$, representing 
the normalization and power-law index;
\item $n_\gamma$ is the number of ionizing 
photons intrinsically emitted per stellar baryon; and 
\item $f_{\rm esc}\left(M_{\rm vir}\right) = \min\left[1, f_{{\rm esc},10} \left(\frac{M_{\rm vir}}{10^{10}{\msol}}\right)^{\alpha_{\rm esc}}\right]$
is the ionizing escape fraction defined as the number ratio of photons reaching 
the IGM to those emitted in the galaxy, and is also assumed to scale with the 
halo mass 
\citep{Ferrara2013MNRAS.431.2826F,Kimm2014,Paardekooper2015MNRAS.451.2544P,Xu2016ApJ...833...84X}.
\end{enumerate}

Inside the $\hii$ regions, we estimate the local, average photoionization rate with 
\citep{Sobacchi2014MNRAS.440.1662S}:
\begin{equation}\label{eq:gamma}
{\bar{\Gamma}_{\rm ion}\left({\bf r},z\right)} = \left(1+z\right)^2 R\sigma_{\rm H}\frac{\alpha_{\rm UVB}}{\beta_{\rm H}+\alpha_{\rm UVB}}\frac{\bar{\rho}_{\rm b}}{m_{\rm p}}{\dot{\bar{n}}_{\rm ion}},
\end{equation}
where $\sigma_{\rm H}=6.3\times10^{-18}{\rm cm}^2$ and $\beta_{\rm H}{\sim}2.75$ are the photoionization cross-section at Lyman limit and its spectral dependence; $\alpha_{\rm UVB}{\sim}5$, $m_{\rm p}$ and $\dot{\bar{n}}_{\rm ion}$ are
the spectral indices of a stellar-driven UV ionizing 
background \citep{Thoul1996ApJ...465..608T}, the mass of a proton, and the local production rate of ionizing photons. Assuming the 
typical star formation time-scale is $t_*H^{-1}\left(z\right)$, with $t_*$ being a free parameter,
we calculate the average 
star formation rate (SFR) of galaxies in halos of a given mass at a given redshift by
\begin{equation}\label{eq:sfr}
{\rm SFR}\left(M_{\rm vir},z\right) = \frac{M_*}{t_* H(z)^{-1}},
\end{equation}
convert it to the non-ionizing UV luminosity via $L_{1500}/{\rm SFR}=8.7\times10^{27}{\rm erg\ s^{-1}Hz^{-1}{\msol}^{-1}yr}$ \citep{Madau2014ARA&A..52..415M},
and estimate $\dot{\bar{n}}_{\rm ion}$ using equation (\ref{eq:n_ion}) with $M_*$ 
being replaced by the SFR.

We follow \citet{Sobacchi2014MNRAS.440.1662S} and estimate the recombination 
rate in each cell with a spatial position and redshift of ({\bf r, $z^\prime$}) as well 
as an overdensity of $\Delta_{\rm cell}$ by
\begin{equation}
\dot{n}_{\rm rec}\left({\bf r},z^\prime \right){=}\int{\rm d}\Delta_{\rm sub}\phi_{\rm sub}\alpha_{\rm B}f_{\rm H} \frac{\bar{\rho}_{\rm b}}{m_{\rm p}} \frac{\Delta_{\rm sub}^2}{\Delta_{\rm cell}} \left(1{-}x_{\rm \hone, sub}\right)^2.
\end{equation}
Here $\Delta_{\rm sub}$, 
$\phi_{\rm sub}\left(z^\prime, \Delta_{\rm sub}| \Delta_{\rm cell}\right)$, $\alpha_{\rm B}$, 
$f_{\rm H}$ and $x_{\hone, {\rm sub}}\left(z^\prime, \Delta_{\rm sub}, T_{\rm g}, \bar{\Gamma}_{\rm ion}\right)$ 
are the sub-grid (unresolved) overdensity, its probability distribution function (PDF; e.g. \citealt{Miralda2000ApJ...530....1M}) within our 
large-scale simulation cell (${\sim}2$Mpc), case-B recombination coefficient evaluated at $10^4{\rm K}$, 
number fraction of hydrogen in the Universe, and the neutral hydrogen fraction of the sub-grid gas element, respectively.
We assume photoionization 
equilibrium in the ionized IGM, accounting for the attenuation of the 
local photoionization rate, $\bar{\Gamma}_{\rm ion}$, according to the radiative transfer 
simulations from \citet{Rahmati2013MNRAS.430.2427R}.
We then compute the cumulative number of recombinations for each cell (see equation \ref{eq:ionization}) by integrating $\dot{n}_{\rm rec}$ from the time the cell was ionized to the redshift of interest.

In summary, our model consists of the following six astrophysical parameters that we sample within the MCMC:
\begin{itemize}
\item $f_{*,10}$, the normalisation of the stellar-to-galactic gas mass relation at $M_{\rm vir}=10^{10}{\msol}$, sampled with a flat prior in log space between $10^{-3}$ and 1;
\item $\alpha_*$, the power-law index of the stellar-to-galactic gas mass relation, sampled with a flat prior between -0.5 and 1;
\item $f_{\rm esc,10}$, the normalisation of the ionizing escape fraction to halo mass relation at $M_{\rm vir}=10^{10}{\msol}$, sampled with a flat prior in log space
between $10^{-3}$ and 1;
\item $\alpha_{\rm esc}$, the power-law index of the  ionising escape fraction to halo mass relation, sampled with a flat prior between -1 and 0.5;
\item $M_{\rm turn}$, the turnover halo mass below which the number density of halos hosting star-forming galaxies drops exponentially, sampled with a flat prior in log space between $10^8$ and $10^{10}{\msol}$;
\item $t_*$, the star-formation timescale as a fraction of the Hubble time, sampled with a flat prior between 0 and 1.
\end{itemize}
These prior ranges are chosen based on the physical meaning of the parameters.  For example, fractions must range from 0 to 1, and we observe galaxies inside halos with masses around $10^{10}{\msol}$ thus setting an upper limit on $M_{\rm turn}$.  More detailed discussion on the parameters and corresponding observational constraints can be found in \citet{Park2019MNRAS.484..933P}.

We stress that our EoR model is both {\it flexible} and {\it physical}; both properties are important for useful inference.  It is {\it flexible} in that it is capable of reproducing (see e.g. \citealt{Park2019MNRAS.484..933P}) the bulk properties and scalings of high-redshift galaxy observations (e.g. \citealt{Bouwens2015a,Bouwens2016,Oesch2018ApJ...855..105O}), 
as well as results from more sophisticated semi-numerical models and hydrodynamic simulations (e.g. \citealt{Mutch2016,Xu2016ApJ...833...84X,Tacchella2018ApJ...868...92T,Behroozi2019MNRAS.488.3143B,Yung2019MNRAS.490.2855Y}). It is {\it physical} in that the equations and parameters have a straightforward interpretation in terms of galaxy evolution, allowing us to set physically-motivated priors for the free parameters.

For computational convenience, we fix the cosmology to the best-fit values of {\it Planck} 2018 {\it TT+lowE}
($\Omega_{\mathrm{m}}, \Omega_{\mathrm{b}}, \Omega_{\mathrm{\Lambda}}, h, \sigma_8$ = 
0.321, 0.04952, 0.679, 0.6688, 0.8118). In practice, one should co-vary astrophysical and cosmological parameters when performing inference. Fixing cosmological parameters effectively assumes that they are mostly constrained by the 
temperature and higher-$\ell$ polarization PS {\color{black}(which are not considered in the 
likelihoods in this work)}, while the $\ell\le29$ E-mode PS constrains the EoR. 
{\color{black} There are less than 5 percent differences ($<$ 1/3 $\sigma$) in the recovered optical 
depth from 
ignoring the 
temperature component in the likelihood.}
However, the most important 
degeneracy affecting the determination of $\tau_e$ is the known ``$A_s\exp(-2\tau_e)$'' 
degeneracy \citep{Planck2016A&A...596A.107P}.  Using the {\it Tanh} model, we show in 
Appendix \ref{sec:appendix} that co-varying $\sigma_8$ following that degeneracy has no 
impact on the reconstructed EoR history.

\subsection{Hyperbolic tangent vs. physically motivated EoR model}\label{subsec:tanhvs21cmfast}

We perform MCMC simulations for the two different EoR models ({\it Tanh} and {\tocm}), both constrained using the low-$\ell$ EE PS. As an additional constraint, we also include the model-independent upper limit on the neutral hydrogen fraction at $z\sim5.9$, $\bar{x}_{\hone}<0.06{+}0.05(1\sigma)$, measured from the dark fraction in QSO spectra \citep{McGreer2015MNRAS.447..499M} and modelling the associated likelihood as a one-sided Gaussian \citep{Greig2017}.

\begin{figure*}
	\begin{minipage}{\textwidth}
		\begin{center}
			\includegraphics[width=0.90\textwidth]{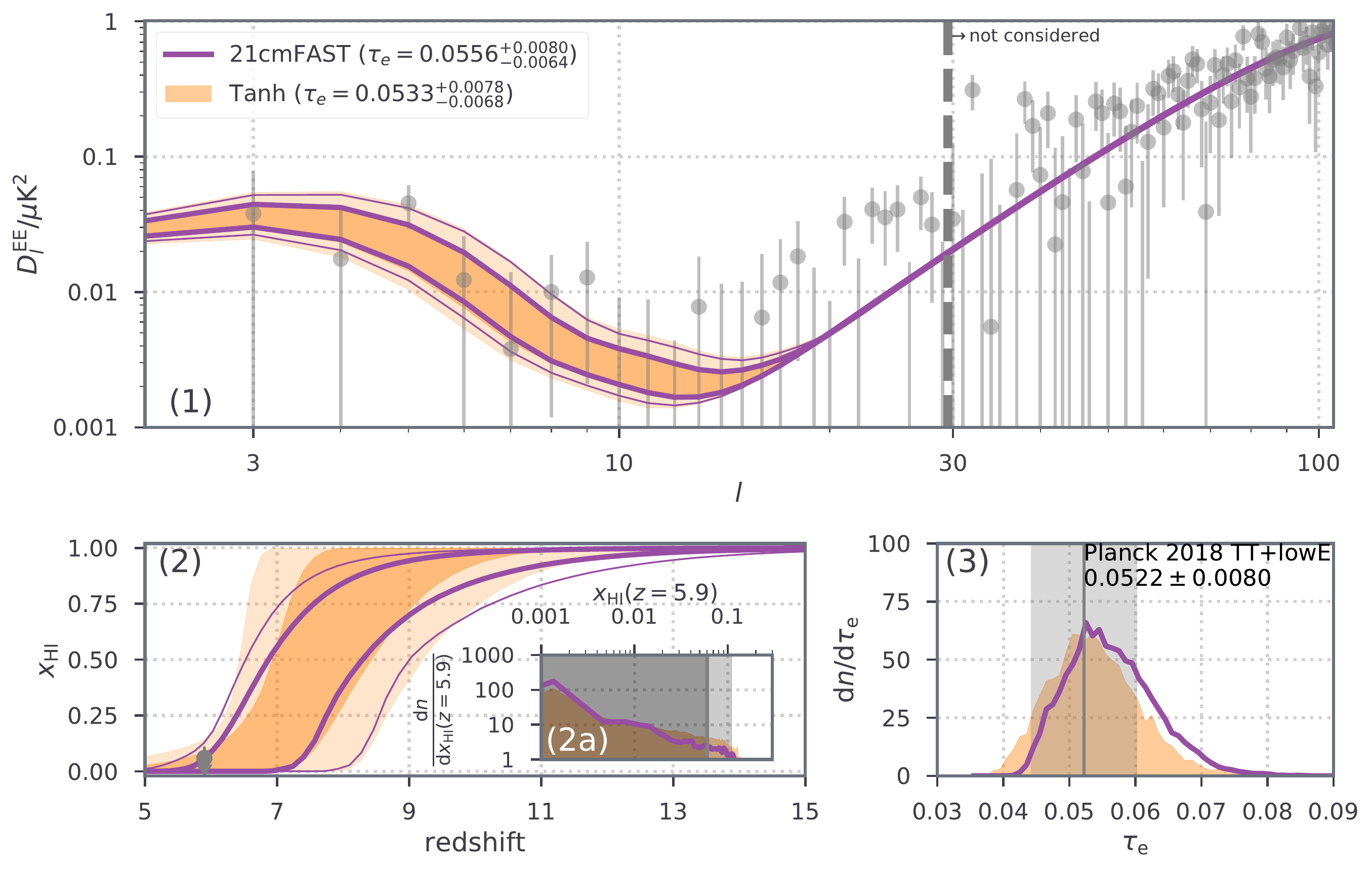}
		\end{center}
	\end{minipage}
	\caption{\label{fig:MCMC_tanh} Posteriors of our astrophysical EoR model ({\it purple}) and the {\it Tanh} EoR model ({\it orange}), recovered from the E-mode polarization PS on large scales ($2\le \ell \le 29$; \citealt{Planck2019arXiv190712875P}; grey circles shown in panel 1) and the upper limit on the neutral hydrogen fraction at $z\sim5.9$ (\citealt{McGreer2015MNRAS.447..499M}; grey shaded regions shown in panel 2a). 
		The 14 to 86 (2.3 to 97.7) percentiles of (1) the E-mode polarization PS and (2) the mean neutral hydrogen fraction evolution in the posteriors are presented using thick (thin) lines or dark (light) shaded regions. PDFs of (2a) the neutral hydrogen fraction at $z=5.9$ and (3) the Thompson optical depth are also shown for the two MCMC results.
		Although the recovered PS and $\tau_e$ are comparable between the two models, the astrophysical model (based on the hierarchical structural growth) results in asymmetric EoR histories with tails towards high redshifts.
	}
\end{figure*}

In Fig. \ref{fig:MCMC_tanh}, we show the resulting posterior for the {\it Tanh} model using orange shaded regions, while the posterior corresponding to the astrophysical model is denoted with purple lines ([14, 86] and [2.3, 97.7] percentiles).  The recovered EE PS, $\tau_e$, and EoR history are shown clockwise from the top panel.

In the top panel, we see that the differences in the recovered PS are negligible between the two models.  Neither model is able to recover the excess power at $20\lsim \ell \lsim 30$. However, the probability-to-exceed of these data as computed in \cite{Planck2019arXiv190712875P} points to a statistical fluctuation rather than a residual systematic error (or new physics not captured by our EoR modelling).

In the bottom right panel, we see that the distributions of $\tau_e$ peak on similar scales, though the shape of the PDFs are qualitatively different.  The recovered median and [14,86] percentiles are comparable (i.e. $\tau_e = 0.0533^{+0.0078}_{-0.0068}$ for {\it Tanh}; $\tau_e = 0.0556^{+0.0080}_{-0.0064}$ for the astrophysical model).  However, we note that the astrophysical model results in an asymmetric PDF of $\tau_e$, with a tail extending towards high values.

The reason for this is apparent looking at the recovered EoR histories in the bottom left panel.  The two distributions are comparable around the midpoint of reionization, where a large fraction of the EE power is imprinted.  However, the astrophysical model (based on the growth of structure) results in asymmetric EoR histories with a tail towards higher redshifts.  For example, at $z=10$, the astrophysical model recovers $\bar{x}_{\hone} \gsim 0.849 (1 \sigma)$, while the {\it Tanh} model recovers $\bar{x}_{\hone} \gsim 0.925 (1 \sigma)$.
Thus, although the preference for a higher median optical depth is reduced in {\it Planck} 2018 compared to the 2015 data (see section \ref{subsubsec:planck}), the {\it natural shape} of the EoR history implied by the growth of structure does result in a (modest) high-redshift tail.

\subsection{Optical depth vs. power spectra}\label{subsec:tauvsps}

In the previous section we discussed how inference from the {\it Planck} 2018 E-mode PS is affected by the choice of EoR models -- {\it Tanh} and astrophysical.  In this section, we only use the astrophysical EoR model, and instead investigate the impact of the choice of likelihood statistic -- using the EE PS directly vs using the $\tau_e$ summary statistic.  In addition to the choice of CMB statistics, we also account for the (model independent) upper limit on the neutral hydrogen fraction at $z\sim5.9$ from the dark fraction in QSO spectra \citep{McGreer2015MNRAS.447..499M} as well as the galaxy UV luminosity functions (LFs) at $z{=}6{-}10$ from \citet{Bouwens2015a,Bouwens2016,Oesch2018ApJ...855..105O}; both of these independent data sets are included as priors in the total likelihood, as described in \citet{Park2019MNRAS.484..933P}.

\begin{figure*}
	\begin{minipage}{\textwidth}
		\begin{center}
			\includegraphics[width=\textwidth]{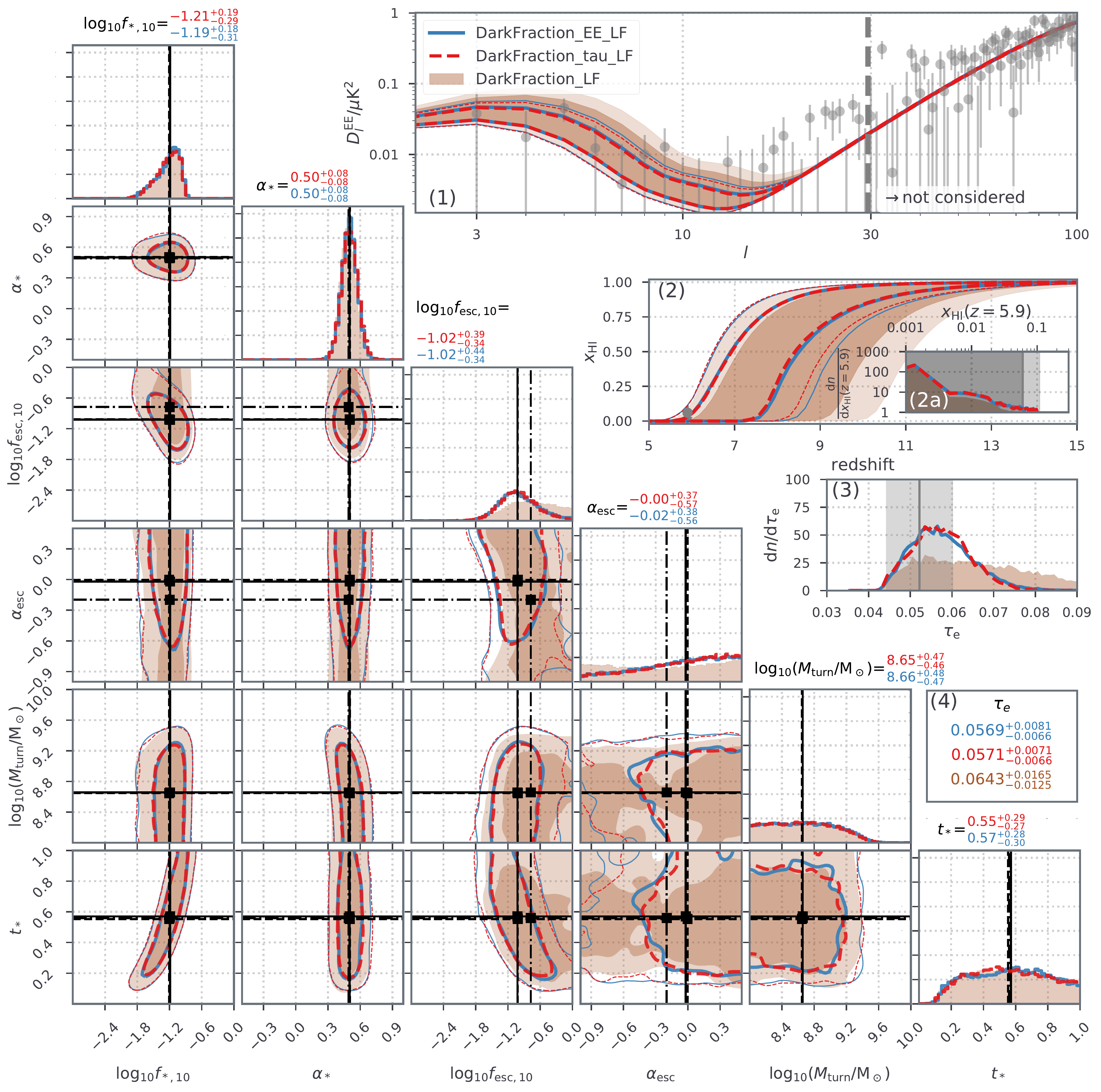}
		\end{center}
	\end{minipage}
	\caption{\label{fig:MCMC}Marginalized posterior distributions of the astrophysical model 
		parameters with different observational constraints:
		(i) \textit{DarkFraction\_LF} (brown shaded regions) uses high-redshift LFs and the QSO dark fraction measurements; 
		(ii) \textit{DarkFraction\_tau\_LF} (red dashed lines) uses high-redshift LFs, the QSO dark fraction measurements and $\tau_e$ derived using a {\it Tanh} EoR model; and
		(iii) \textit{DarkFraction\_EE\_LF} (blue solid lines) uses high-redshift LFs, the QSO dark fraction measurements, and the low-$\ell$ EE PS.
		The 2D distributions correspond to 68th (dark regions or thick lines) and 95th (light regions or thin lines) percentiles.
		The upper-right three sub-panels 
		present the [14, 86] (and [2.3, 97.7]) percentiles of (1) the E-mode 
		polarization PS and (2) evolution of the mean neutral hydrogen fraction 
		($\bar{x}_{\hone}$), as well as the PDFs of (2a) $x_{\hone}$ 
		at $z{=}5.9$ and (3) the Thompson scattering optical depth.
		The median and [14, 86] percentiles of the inferred 
		optical depth are presented in the lower right box (4).
		Observational constraints are indicated in 
		grey. Overall, we see {\it insignificant difference} in the posteriors 
		between
		\textit{DarkFraction\_tau\_LF} and \textit{DarkFraction\_EE\_LF}, indicating a negligible bias in inference when using the $\tau_e$ summary statistic generated with a different EoR basis set (instead of directly forward-modelling the EE PS).}
\end{figure*}

In Fig. \ref{fig:MCMC}, we show the marginalized posteriors for the astrophysical parameters (panels in {\it lower left corner}), together with the corresponding: (1) EE PS; (2) EoR history; and (3) $\tau_e$  ({\it upper right}).  We run three different MCMC simulations corresponding to different combinations of observational data sets used for the likelihood:
\begin{enumerate}
	\item \textit{DarkFraction\_LF}, including high-redshift LFs and the QSO dark fraction upper limit on $\bar{x}_{\hone}(z=5.9)$. This run does not consider any CMB observations;
     \item \textit{DarkFraction\_tau\_LF}, based on \textit{DarkFraction\_LF}, but including an additional constraint on $\tau_e = 0.0522\pm0.0080$, which is taken from the {\it TT+lowE} reconstruction in {\it Planck} 2018 \citep{Planck2018arXiv180706209P} and was generated using a {\it Tanh} basis set; and
     \item \textit{DarkFraction\_EE\_LF}, based on \textit{DarkFraction\_LF},  but also forward-modelling the low-$\ell$ E-mode PS using \textsc{class} \citep{Lesgourgues2011arXiv1104.2932L} and includes {\color{black}the low-$\ell$ EE likelihood from {\it Planck} 2018} \citep{Planck2019arXiv190712875P}. These are our ``flagship'' constraints. 
\end{enumerate}

Looking at the posterior of \textit{DarkFraction\_LF}, we see that some of our astrophysical parameters are already constrained by galaxy and QSO observations.  As pointed out in \citet{Park2019MNRAS.484..933P} (see also e.g. \citealt{Tacchella2018ApJ...868...92T,Behroozi2019MNRAS.488.3143B,Yung2019MNRAS.490.2855Y}), the observed high-redshift UV LFs already constrain the SFR-to-halo mass relation to within factors of ${\sim}2$ (i.e. $f_{*, 10}/t_\ast$ and $\alpha_\ast$) and provide an upper limit on the characteristic halo mass below which the galaxy occupancy fraction starts decreasing ($M_{\rm turn}\lsim 10^{9.5} {\msol}$; 2$\sigma$).
Additionally, the dark fraction measurement of QSO spectra sets a lower limit for the ionizing escape fraction normalization ($f_{\rm esc, 10} > 0.02$; 2$\sigma$), requiring the bulk of reionization to occur before $z\sim6$ (panel 2) and setting a lower limit on $\tau_e$ (panel 3). However, without the CMB, the early
stages of reionization are unconstrained (c.f. \citealt{Greig2017}).  This is evident from the broad range of EoR histories allowed beyond $z\gtrsim8$
(panel 2), as well as the broad distributions of $\tau_e$ (panel 3) and the PS (panel 1).  These early EoR models generally correspond to the high $f_{\rm esc}$ + low $\alpha_{\rm esc}$ corner of astrophysical parameter space.\footnote{We note that if our prior ranges were extended even further, reionization would be allowed at even higher redshifts.  However, an important benefit of using an astrophysical EoR model is that it allows us to place physically-motivated priors on the parameters.  For example, an ionizing escape fraction cannot be higher than unity, nor can star formation occur efficiently inside halos whose virial temperature is smaller than available gas cooling channels.  This is not the case for non-physical or so-called model-independent constraints, for which it can be difficult to choose reasonable priors on the model parameters.}

Including CMB observations rules out early reionizing models.  Both \textit{DarkFraction\_tau\_LF} and \textit{DarkFraction\_EE\_LF} posteriors disfavour the high $f_{\rm esc}$ and low $\alpha_{\rm esc}$ corner of parameter space. The EE PS, EoR history and $\tau_e$ distributions all shrink.

Comparing the  \textit{DarkFraction\_tau\_LF} and \textit{DarkFraction\_EE\_LF} posteriors in Fig. \ref{fig:MCMC} quantifies the bias of using the $\tau_e$ summary statistic, generated with a different EoR model, instead of directly forward-modelling the CMB PS.  There is a small difference in the EE PS with \textit{DarkFraction\_EE\_LF} allowing for a slightly earlier EoR. In general however the two posteriors are nearly identical.  This indicates that, with the {\it Planck} 2018 data, the bias in using $\tau_e$ for the likelihood instead of the EE PS directly is negligible.

\begin{figure}
	\begin{minipage}{\columnwidth}
		\begin{center}
			\includegraphics[width=\textwidth]{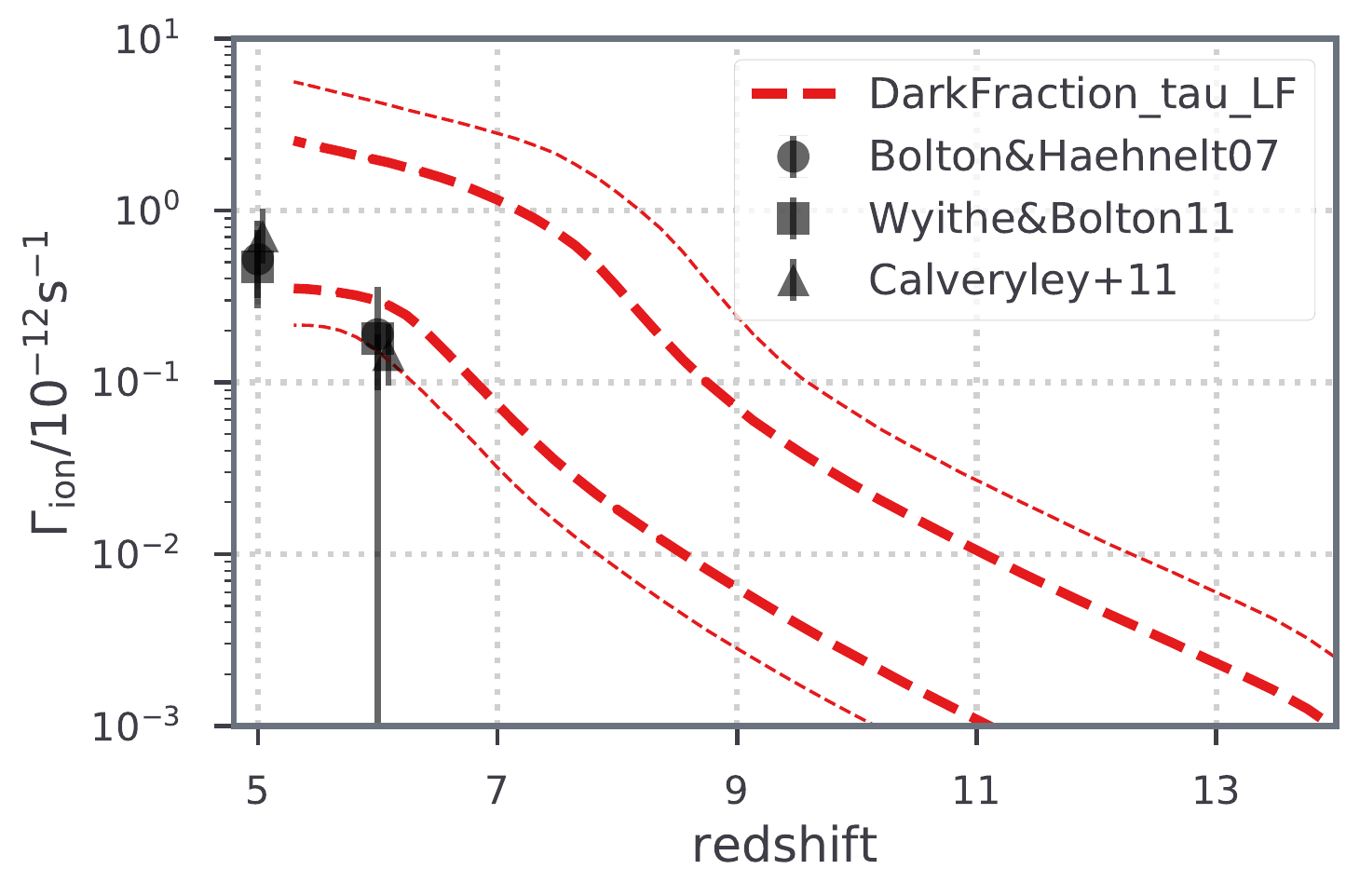}
		\end{center}
	\end{minipage}
	\caption{\label{fig:gamma}{\color{black}The [14, 86] (and [2.3, 97.7]) percentiles of the photonionization rate from models used in the {\it DarkFraction\_tau\_LF} posterior in Fig. \ref{fig:MCMC}. Observational constraints \citep{Bolton2007MNRAS.382..325B,Wyithe2011MNRAS.412.1926W,Calverley2011MNRAS.412.2543C} are indicated in grey.}
	}
\end{figure}

{\color{black}
Finally, Fig. \ref{fig:gamma} presents the constrained photonionization rate ($\Gamma_{\rm ion}$; see equation \ref{eq:gamma}).  This serves to further illustrate how a physical model allows us to predict additional IGM properties.  We only show {\it DarkFraction\_tau\_LF}, as {\it DarkFraction\_EE\_LF} is nearly identical, while not including the CMB observations allows $\Gamma_{\rm ion}$ to become unrealistically large.
The rise in $\Gamma_{\rm ion}$ with redshift is determined by the formation of structure in our galaxy model, with the ``flattening'' seen at $z\lsim$6--7 being due to photo-heating suppression of gas accretion onto galaxies following reionization
(e.g. \citealt{Sobacchi2013MNRAS.432L..51S}).
We see that while our prediction is consistent with the measured UV ionizing background at $z\sim 5$ to 6 \citep{Bolton2007MNRAS.382..325B,Calverley2011MNRAS.412.2543C,Wyithe2011MNRAS.412.1926W}, observational data lies on the lower boundary.
Additionally including $z\sim5$--6 Ly$\alpha$ transmission statistics constrains the upper envelope of $\Gamma_{\rm ion}$ significantly, though the results are more model dependent than those presented here (Qin et al. in prep).
}

We should caution however that our findings here are valid in the context of a flat $\Lambda$CDM Universe. An important follow-up to this study will be to generalize these trends to alternative cosmologies (e.g. \citealt{Paoletti2020arXiv200512222P}).
It is possible that in some exotic cosmologies, the correlation between parameters describing exotic physics and reionization exist at the level of $\tau_e$ but are broken once the full power of the CMB PS is included, rendering the necessity of joint analysis of reionization data and CMB observations.

\section{Conclusions}\label{sec:conclusion}

In this work, we develop an interface between {\tocm} and \textsc{class}, allowing us to forward-model the large-scale E-mode polarization power spectra inside the {\cmmc} sampler.  With this setup, we study how the choice of (i) EoR model (astrophysical vs. a {\it Tanh}), and (ii) likelihood statistic ($\tau_e$ vs EE PS), impacts EoR parameter inference.

The marginalized posteriors of $\tau_e$ for the {\it Planck} 2018 data \citep{Planck2019arXiv190712875P} are fairly insensitive to the parametrization of the EoR history: {\it Tanh} vs a galaxy model.  This is contrary to claims based on the earlier 2015 release (e.g. \citealt{Miranda2017MNRAS.467.4050M,Planck2018arXiv180706209P,Millea2018A&A...617A..96M,Hazra19}). However, the galaxy model (based on standard hierarchical growth of structure) results in asymmetric EoR histories, with the early stages extending to higher redshifts (see also \citealt{Choudhury2006MNRAS.371L..55C,Wise2014,Price2016,Qin2017a,Gorce2018A&A...616A.113G,Kulkarni2018}).  As a result, the lower limit on the volume averaged neutral hydrogen fraction at $z=10$ changes from $\bar{x}_{\hone} \gsim 0.93 (1 \sigma)$ using the {\it Tanh} model to $\bar{x}_{\hone} \gsim 0.85 (1 \sigma)$ using an astrophysical model.

Using our galaxy EoR model, we quantify the bias in inference when the likelihood is computed from the $\tau_e$ summary statistic, compared with directly using the
the measured E-mode polarization PS. We perform MCMC simulations 
taking into account the current observational constraints from high-redshift 
galaxy UV LFs \citep{Bouwens2015a,Bouwens2016,Oesch2018ApJ...855..105O}
and the model-independent constraints from the dark fraction in QSO spectra \citep{McGreer2015MNRAS.447..499M}.
Additionally including either CMB statistic helps constrain the posterior, ruling out models that have a high escape fraction for faint galaxies and hence an early reionization.  However, the difference between using $\tau_e$ for the likelihood, compared with the EE PS, is negligible.
Our flagship constraints, based on the QSO dark fraction + UV LFs + EE PS, result in an optical depth of
$\tau_e = 0.0569_{-0.0066}^{+0.0081} (1\sigma)$, with asymmetric EoR histories.

\section*{Acknowledgements}
The authors thank V. Miranda for providing data from \citet{Miranda2017MNRAS.467.4050M}.
This work was supported by the European Research Council (ERC) under the European Union’s 
Horizon 2020 research and innovation programme (AIDA -- \#638809). The results presented here
reflect the authors’ views; the ERC is not responsible for their use. Parts of this research 
were supported by the Australian Research Council Centre of Excellence for All Sky 
Astrophysics in 3 Dimensions (ASTRO 3D), through project \#CE170100013.

\section*{Data availability}
The data underlying this article will be shared on reasonable request to the corresponding author.

\bibliographystyle{mn2e}
\bibliography{reference}

\appendix

\section{reconstructed EoR when covarying $\sigma_8$}\label{sec:appendix} 

\begin{figure*}
	\begin{minipage}{\textwidth}
		\begin{center}
			\includegraphics[width=0.90\textwidth]{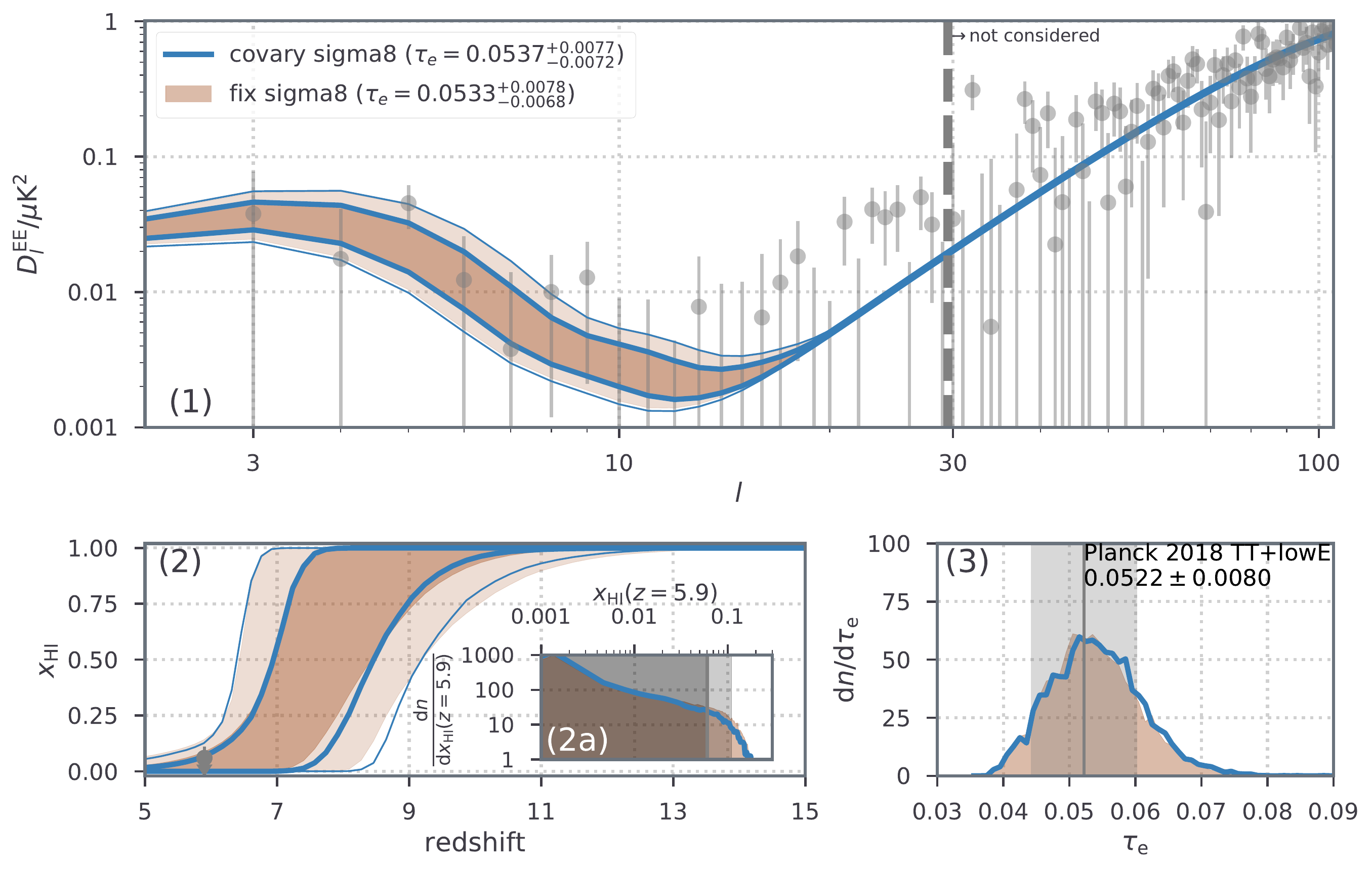}
		\end{center}
	\end{minipage}
	\caption{\label{fig:MCMC_tanh_covarysigma8} Same as Fig. \ref{fig:MCMC_tanh}, but for the {\it Tanh} EoR model with ({\it blue}) and without co-varying $\sigma_8$ ({\it orange}).  For the posterior shown in blue, we put a prior on the $\sigma_8$ -- $\tau_e$ relation inferred from the high-$\ell$ data (see Fig. 42 in \citealt{Planck2016A&A...596A.107P}).   The difference between the two results is negligible.
	}
\end{figure*}

It is well known that there exists a strong degeneracy between the optical depth to reionization $\tau_e$ and the amplitude of the primordial power spectrum $A_s$ in the high-$\ell$ ($\ell\gtrsim30$) TT,TE,EE power spectrum, such that the parameter combination well-constrained by these data is $A_s\exp(-2\tau_e)$ \citep{Planck2016A&A...596A.107P}. Once the physical densities $\Omega_{\rm m}h^2$, $\Omega_{\rm b}h^2$ and the tilt of the primordial power spectrum $n_s$ are fixed, there is a direct correspondence between $A_s$ and $\sigma_8$, and hence $\sigma_8$ inherits this degeneracy with $\tau_e$. As a result, one might worry that fixing $\sigma_8$, as we have done for computational convenience, would over-constrain the optical depth and the EoR reconstructed from the low-$\ell$ PS.

To test this, in Fig. \ref{fig:MCMC_tanh_covarysigma8} we plot the analogous quantities from Fig. \ref{fig:MCMC_tanh}, using the {\it Tanh} parametrization, which is much less time-consuming than {\tocm}. In orange, we show the same model as in Fig. \ref{fig:MCMC_tanh}, generated by fixing $\sigma_8$.  In blue we show the posterior of the {\it Tanh} model, but also allowing $\sigma_8$ to co-vary with a prior on the $\sigma$ -- $\tau_e$ relation inherited from the high-multipole data (see Fig. 42 and discussion in \citealt{Planck2016A&A...596A.107P}).
The fact that the orange and blue posteriors are virtually indistinguishable suggests that our conclusions are unaffected by our choice of fixing cosmological parameters while performing inference.

\bsp
\label{lastpage}
\end{document}